\documentclass{article}
\usepackage{spconf,graphicx}
\usepackage{amsmath,epsfig,amssymb,bm,setspace,multirow,pifont}
\usepackage{algorithm,tabularx}
\usepackage[noend]{algpseudocode}
\usepackage{stackengine}

\makeatletter
\newcommand{\multiline}[1]{%
  \begin{tabularx}{\dimexpr\linewidth-\ALG@thistlm}[t]{@{}X@{}}
    #1
  \end{tabularx}
}
\makeatother

\newcommand{\argmin}{\operatornamewithlimits{argmin}}

\title{Federated Acoustic Modeling For Automatic Speech Recognition}
\name{Xiaodong Cui, Songtao Lu and Brian Kingsbury}
\address{IBM Research AI \\ IBM T. J. Watson Research Center, Yorktown Heights, NY 10598, USA}

\begin{document}
\ninept

\maketitle
\begin{abstract}
Data privacy and protection is a crucial issue for any automatic speech recognition (ASR) service provider when dealing with clients. In this paper, we investigate federated acoustic modeling using data from multiple clients. A client's data is stored on a local data server and the clients communicate only model parameters with a central server, and not their data. The communication happens infrequently to reduce the communication cost. To mitigate the non-iid issue, client adaptive federated training (CAFT) is proposed to canonicalize data across clients. The experiments are carried out on 1,150 hours of speech data from multiple domains. Hybrid LSTM acoustic models are trained via federated learning and their performance is compared to traditional centralized acoustic model training. The experimental results demonstrate the effectiveness of the proposed federated acoustic modeling strategy. We also show that CAFT can further improve the performance of the federated acoustic model.
\end{abstract}
\section{Introduction}
\label{sec:intro}

Large-scale acoustic modeling for automatic speech recognition (ASR) usually relies on speech data from multiple sources. For industrial ASR products, oftentimes the acoustic models are customized for one or more target clients. The traditional approach to training acoustic models aggregates the speech data from the clients, along with other public and/or internal speech data if necessary. The training is typically carried out in a centralized fashion on the servers of the ASR service provider. With data privacy and protection becoming a crucial issue in information technology \cite{gdpr}, most clients will require that their data stay on-premises, precluding its release to the provider for training. A new approach to training the models under this circumstance is needed. In this paper, we investigate a federated acoustic modeling strategy where the acoustic model is trained collaboratively among the clients with each client having their data locally stored. The clients only exchange their local model updates with the central server at the service provider. This model exchange takes place at a minimal frequency to reduce the communication cost.\footnote{\copyright 2021 IEEE. Personal use of this material is permitted. Permission from IEEE must be obtained for all other uses, in any current or future media, including reprinting/republishing this material for advertising or promotional purposes, creating new collective works, for resale or redistribution to servers or lists, or reuse of any copyrighted component of this work in other works.}

Federated learning (FL) \cite{Konecny_FL,Kairouz_FL,McMahan_FL} has been widely used in applications such as healthcare and finance where data privacy is a constraint. In the setting of FL, multiple entities collaborate with each other to optimize a machine learning problem under the orchestration of a central server. A global model is learned with each client keeping its data private, in local storage. Even though FL was proposed initially in scenarios with a huge number of mobile or edge devices \cite{Hard_FLKeyboard}, it was later generalized to a much broader spectrum of applications. In \cite{Kairouz_FL}, the initial FL setting with an emphasis on a large number of devices, each with a relatively small amount of data, is referred to as ``cross-device'' FL, while the setting that we are about to use in this work for federated acoustic modeling is referred to as ``cross-silo'' FL. Cross-silo FL deals with a federation of a few data providers each with a relatively large amount of siloed data. In the speech community, FL related efforts have also been reported recently. In \cite{Leroy_FLKWS}, federated learning was used to train an embedded wake word detector on crowdsourced speech. In \cite{Hard_FLKWS}, various federated averaging schemes and data augmentation techniques have been studied to improve keyword spotting models with data not independent and identically distributed (iid) at the edge. An interactive system was built in \cite{Tan_FLASR} to demonstrate how FL can help transfer learning on acoustic models using edge device data. In \cite{Dimitriadis_FL}, a federated transfer learning platform is introduced with improved performance using enhanced federated averaging via hierarchical optimization and gradient selection.

\begin{figure}[htb]
  \centering
  \centerline{\includegraphics[width=4.5cm, height=3.9cm]{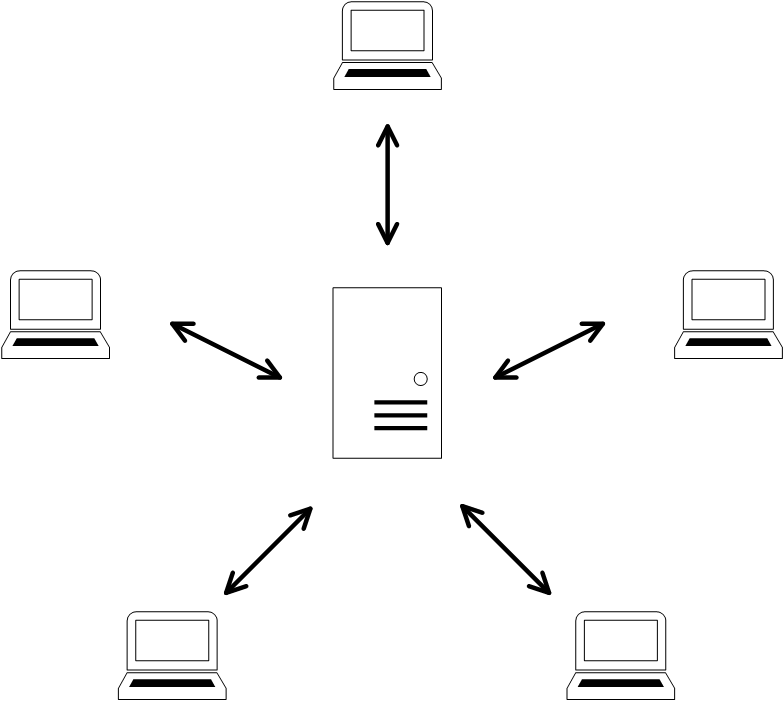}}
  \caption{Federated acoustic modeling with multiple clients.}\label{fig:flam}
\end{figure}

In this paper, we introduce a cross-silo FL framework for joint acoustic modeling with heterogeneous client data from multiple clients. Its configuration is shown in Fig.\ref{fig:flam}. Each client has a local server for data storage and computation. The clients only communicate with the central server of the service provider. Model parameters, not raw data, are exchanged between the clients and the central server in each round of communication. Local model parameters are uploaded to the central server and aggregated into a global model that is then transmitted back to each client. The communication is synchronous and takes place at a minimal frequency to reduce the communication cost. Since client data may come from distinct domains with unbalanced amounts, a fundamental issue with federated acoustic modeling in the real world is dealing with non-iid data. In this work, we propose a client adaptive federated training (CAFT) strategy to mitigate data heterogeneity. Experiments are conducted on 1,150 hours of speech data from multiple domains including public data, internal data, and real-world client data. We compare the performance of the federated strategy under various settings and also compare it with the traditional centralized training.

The remainder of the paper is organized as follows. Section \ref{sec:prob} gives the mathematical formulation the federated acoustic modeling. Section \ref{sec:fam} and Section \ref{sec:catfam} present the algorithms and implementation of federated training and client adaptive federated training of acoustic models. Experimental results are reported in Section \ref{sec:exp} followed by a discussion in Section \ref{sec:dis}. Finally, we conclude the paper with a summary in Section \ref{sec:sum}.

\section{Problem Formulation}
\label{sec:prob}

Suppose there are $L$ clients. The data of client $i$, $i=1,\cdots,L$, consists of $n_{i}$ samples which can only be accessed locally. The data follows a client-dependent distribution $\mathcal{D}_{i}$. Each client is associated with a local risk function
\begin{align}
    \mathcal{L}_{i}(w)\triangleq \mathbb{E}_{x \thicksim \mathcal{D}_{i}}[f(w,x)]  \label{eqn:lloss}
\end{align}
where $w$ is the parameters, $x \thicksim \mathcal{D}_{i}$ the data samples from distribution $\mathcal{D}_{i}$ and $f(w,x)$ the loss function.

Federated modeling optimizes the following risk function over $L$ clients
\begin{align}
    \min _{w} \  \mathcal{L}(w) & = \sum_{i=1}^{L}p_{i}\mathcal{L}_{i}(w) = \sum_{i=1}^{L}p_{i}\mathbb{E}_{x \thicksim \mathcal{D}_{i}}[f(w,x)] \label{eqn:gloss}
\end{align}
where $p_{i}\!>\!0$ and $\sum_{i}p_{i}\!=\!1$ are weights on the local risk functions. It is typical to choose $p_{i}=\frac{1}{L}$ to make the clients equally contributed.

In conventional distributed training \cite{Cui_DistASRSPM,Zhang_ADPSGDSWB,Chen_BMUF,Huang_RandomBMUF}, data from multiple sources are first mixed and then distributed to learners. Each learner has equal access to the mixed data and therefore local distribution across learners is iid. In federated learning, however, the data from different sources can not be mixed and hence the local data distribution is non-iid, which is different from the conventional distributed training.  This is a fundamental issue in federated learning. In addition, the amounts of data from the clients could be unbalanced. As a result, the weights in the global loss function in Eq.\ref{eqn:gloss} are sometimes set to $p_{i}=\frac{n_{i}}{n}$ where $n=\sum_{i=1}^{L}n_{i}$ to make the loss function of each client proportional to their amounts of data.

\section{Federated Acoustic Modeling}
\label{sec:fam}

The recipe for cross-silo federated training of acoustic models is given in Algorithm \ref{alg:fam}.

The central server of the service provider coordinates the distributed training among clients. It starts the training by sending a globally initialized model to the clients and aggregates the locally updated models from all clients for a total of $T$ rounds of communication. In each communication round, clients receive the global model from the central server and update it locally using their own local data before sending it back to the central server. This process is carried out in parallel with synchronization. The central server will wait until models from all clients are in place to update the global model. The local model update can be realized by any optimizer. In this work, stochastic gradient descent (SGD) with momentum is used. The local client data is evenly divided into $T$ chunks and one chunk of data after randomization is used for multi-step mini-batch SGD with a batch size $B$ in each communication round:
\begin{align}
  w_{k+1} =  w_{k} - \alpha \sum_{j=1}^{B}\nabla f(w_{k},x_{j}).
\end{align}

The global model update is conducted by federated averaging (FedAvg) \cite{Konecny_FL}:
\begin{align}
    w_{t+1} = w_{t} - \eta \sum_{i=1}^{L}p_{i}\left(w_{t}-w^{(i)}_{t+1}\right)
\end{align}
Note that when the global learning rate $\eta=1$, FedAvg is equivalent to a simple model averaging strategy
\begin{align}
    w_{t+1} = \sum_{i=1}^{L}p_{i}w^{(i)}_{t+1}
\end{align}
which is analogous to the $K$-step averaging SGD \cite{Zhou_Kstep} when $p_{i}=\frac{1}{L}$. However, in this case the steps employed in each client are different and the data distribution is non-iid.

\begin{algorithm}
\caption{Federated acoustic modeling}\label{alg:fam}
\begin{algorithmic}\smallskip
\State $L$ $\leftarrow$ total number of clients;
\State $M$ $\leftarrow$ total number of epochs;
\State $T$ $\leftarrow$ total rounds of communication;
\State $B$ $\leftarrow$ batch size;
\State $\alpha$ $\leftarrow$ local learning rate;
\State $\eta$ $\leftarrow$ global learning rate;
\medskip
\Function{\textbf{Server}}{}
    \State Initialize model $w$;
    \For  {\ $m \leftarrow 1, \cdots, M$ \ }
        \State copy model from last epoch
        \For  {\ $t \leftarrow 1, \cdots, T$ \ }
            \State Send model $w_{t}$ to clients
            \For {\ $i \leftarrow 1, \cdots, L$ \ } \Comment{running in parallel}
                  \State $w^{(i)}_{t+1}$ $\leftarrow$ \textbf{Client}($i$, $t$)
            \EndFor
            \State $w_{t+1} \leftarrow w_{t} - \eta \sum_{i=1}^{L}p_{i}\left(w_{t}-w^{(i)}_{t+1}\right)$
        \EndFor
    \EndFor
\EndFunction
\medskip
\Function{\textbf{Client}}{$i$, $t$}
    \State Receive model $w_{t}$ from server;
    \State \multiline{Get the samples $S^{(i)}_{t}$ for the communication round where $|S^{(i)}_{t}|= \frac{n_{i}}{T}$;}
    \State Get the number of SGD steps $K_{i} = \frac{|S^{(i)}_{t}|}{B}$;
    \For  {\ $k \leftarrow 1, \cdots, K_{i}$ \ }
        \State  $w^{(i)}_{t,k+1} \leftarrow w^{(i)}_{t,k} - \alpha \sum_{j=1}^{B}\nabla f(w^{(i)}_{t,k},x_{j}), \ \ x_{j} \thicksim S^{(i)}_{t}$;
    \EndFor
    \State Return model $w^{(i)}_{t+1}=w^{(i)}_{t,K_{i}}$ to server;
\EndFunction
\end{algorithmic}
\end{algorithm}

\section{Client Adaptive Federated Training}
\label{sec:catfam}

The data provided by clients may come from different domains with different distributions. Since the clients update their models locally and do not directly communicate with each other, this non-iid issue has posed a major challenge in FL. In ASR, canonicalization has been widely used in dealing with data heterogeneity, for example that caused by different speakers and environments \cite{Gales_SAT,Gales_CMLLR}. In this work, we introduce CAFT to mitigate data heterogeneity across clients, extending speaker or cluster adaptive training to federated learning. We introduce a transform $\mathcal{F}_{i}$ of the data $x$ to each client $i$. The transform is estimated to minimize the local risk function given the global model $w$
\begin{align}
       \hat{\mathcal{F}_{i}} & = \argmin_{\mathcal{F}_{i}} \mathcal{L}\left(w, \mathcal{F}_{i}\right) \nonumber \\
       & \triangleq \argmin_{\mathcal{F}_{i}} \mathbb{E}_{x \thicksim \mathcal{D}_{i}}[f(w,\mathcal{F}_{i}(x))]
\end{align}
After the estimation, the transform $\mathcal{F}_{i}$ is fixed and a local model $w^{(i)}$ is updated on the transformed data $\mathcal{F}_{i}(x)$. In each round, $\mathcal{F}_{i}$ and $w^{(i)}$ are alternately optimized. Since $\mathcal{F}_{i}$ is estimated against the global model $w$, the transformed data $\mathcal{F}_{i}(x)$ is expected to be more homogeneous. In this work, the transform $\mathcal{F}_{i}$ is chosen to be an affine transform
\begin{align}
      \mathcal{F}_{i}(x) = A_{i}x+b_{i}, \ \ \ \  x \thicksim \mathcal{D}_{i}
\end{align}

The recipe for CAFT is given in Algorithm \ref{alg:catfam}.

\begin{algorithm}
\caption{Client adaptive federated training}\label{alg:catfam}
\begin{algorithmic}\smallskip
\State $L$ $\leftarrow$ total number of clients;
\State $M$ $\leftarrow$ total number of epochs;
\State $T$ $\leftarrow$ total rounds of communication;
\State $B$ $\leftarrow$ batch size;
\State $\alpha$ $\leftarrow$ local learning rate;
\State $\eta$ $\leftarrow$ global learning rate;
\medskip
\Function{\textbf{Server}}{}
    \State Initialize model $w$;
    \For  {\ $m \leftarrow 1, \cdots, M$ \ }
        \State copy model from last epoch
        \For  {\ $t \leftarrow 1, \cdots, T$ \ }
            \State Send model $w_{t}$ to clients
            \For {\ $i \leftarrow 1, \cdots, L$ \ } \Comment{running in parallel}
                  \State $w^{(i)}_{t+1}$ $\leftarrow$ \textbf{CATClient}($i$, $t$)
            \EndFor
            \State $w_{t+1} \leftarrow w_{t} - \eta \sum_{i=1}^{L}p_{i}\left(w_{t}-w^{(i)}_{t+1}\right)$
        \EndFor
    \EndFor
\EndFunction
\medskip
\Function{\textbf{CATClient}}{$i$, $t$}
    \State Receive model $w_{t}$ from server;
    \State \multiline{Get the samples $S^{(i)}_{t}$  for the communication round where $|S^{(i)}_{t}|= \frac{n_{i}}{T}$;}
    \State Get the number of SGD steps $K_{i} = \frac{|S^{(i)}_{t}|}{B}$;
    \State Estimate client-dependent transform $\mathcal{F}_{i}$ given $w_{t}$;
    \For  {\ $k \leftarrow 1, \cdots, K_{i}$ \ }
        \State  $w^{(i)}_{t,k+1} \leftarrow w^{(i)}_{t,k} - \alpha \sum_{j=1}^{B}\nabla f(w^{(i)}_{t,k},\mathcal{F}(x_{j})), \ \ x_{j} \thicksim S^{(i)}_{t}$;
    \EndFor
    \State Return model $w^{(i)}_{t+1}=w^{(i)}_{t,K_{i}}$ to server;
\EndFunction
\end{algorithmic}
\end{algorithm}

\section{Experimental Results}
\label{sec:exp}

Experiments are conducted on 1,150 hours of speech data from five sources. It consists of 420 hours of Broadcast News (BN) data, 450 hours of internal dictation data, 100 hours of internal meeting data, 140 hours of hospitality (travel and hotel reservation) data and 40 hours of accented data, respectively. It represents a good coverage of public data (BN), internal data (dictation and meeting) and real-world client data (hospitality and accented). Each data source is treated as a client, giving rise to five clients in total. The data is wideband speech with a 16KHz sampling rate. The acoustic models are evaluated on four test sets. To make the evaluation extensive yet controlled, the four selected test sets are taken from public data and real-world client data, as described in Table \ref{tab:testsets}. The decoding vocabulary comprises 260K words and the language model (LM) is a 4-gram LM with 200M n-grams and modified Kneser-Ney smoothing. The LM training data is selected from a broad variety of sources.

\begin{table}[]
    \centering
    \begin{tabular}{l|l|r}
           Test Set  &   Description                       &  Hours   \\ \hline\hline
             S1      & dev04f test set from Broadcast News &  2.21    \\
             S2      & hospitality speech                  &  0.34    \\
             S3      & Asian-accented speech               &  2.41    \\
             S4      & Latin-accented speech              &  3.12    \\
    \end{tabular}
    \caption{Test sets used for evaluation.} \label{tab:testsets}
\end{table}

The acoustic model is a hybrid Long Short-Term Memory (LSTM) network with 5 bi-directional layers. Each layer has 512 cells with 256 cells in each direction. A linear projection layer with 256 hidden units is inserted between the topmost LSTM layer and the softmax layer consisting of 9,300 output units. These 9,300 units correspond to context-dependent hidden Markov model (HMM) states. The LSTM is unrolled over 21 frames and trained with non-overlapping feature subsequences of that length. The feature input is 40 dimensional log-Mel features with delta and double delta features. The total input dimensionality is 120.

Table \ref{tab:fl} shows the word error rates (WERs) of acoustic models trained by the traditional centralized training and federated training under various hyper-parameters such as global learning rates and communication rounds. The baseline is considered an oracle model as it is trained by mixing all the training data such that training is carried out on iid data. It is trained using SGD with a learning rate of 0.1, a momentum of 0.9, and a batch size of 256. The learning rate anneals by $\frac{1}{\sqrt{2}}$ after the 10th epoch and the training finishes in 20 epochs. In federated training, the optimization of the local models for each client follows a recipe similar to the baseline except the training uses local data. The SGD optimizer employs a learning rate of 0.2, a momentum of 0.9, and a batch size 128. All clients are equally weighted (i.e. $p_{i}\!=\!0.2$). When $\eta\!=\!1.0$, FedAvg amounts to a simple model average. Setting the number of communication rounds involves a tradeoff: to be communication efficient, one wants to minimize the number of communication rounds, but this may degrade the performance of the final model. From the table, 20 rounds of communication gives the best WERs, while more rounds of model averaging may slightly hurt the performance. The best WERs are achieved  after 20 rounds of communication when the global learning rate is 0.95.

\begin{table}[tbh]
\centering
\begin{tabular}{c c | c c c c c} \hline
      $\eta$    &    $T$         &    S1   &   S2   &   S3   &   S4   &   avg   \\ \hline\hline
   \multicolumn{2}{c|}{\textbf{baseline}} & \textbf{13.2}  &  \textbf{23.0}  &  \textbf{13.8}  &  \textbf{13.3}  &  \textbf{15.83}  \\ \hline
       1.0      &     10         &   14.2  &  24.2  &  15.4  &  14.7  &  17.13  \\
      \textbf{1.0}      &    \textbf{20}         & \textbf{14.1}  &  \textbf{23.2}  &  \textbf{15.0}  &  \textbf{13.9}  &  \textbf{16.55}  \\
       1.0      &     30         &   13.8  &  25.4  &  14.2  &  13.7  &  16.78  \\
       1.0      &     40         &   13.9  &  23.9  &  15.2  &  14.3  &  16.83  \\ \hline
       0.80     &     20         &   14.2  &  22.8  &  15.6  &  14.4  &  16.75  \\
       0.90     &     20         &   14.2  &  23.3  &  14.8  &  13.9  &  16.55  \\
      \textbf{0.95}     &  \textbf{20}    &  \textbf{14.1}  &  \textbf{22.9}  &  \textbf{14.7}  &  \textbf{13.6}  &  \textbf{16.33}  \\
       0.98     &     20         &   14.5  &  23.5  &  14.3  &  13.8  &  16.53  \\ \hline
\end{tabular}
\caption{WERs of baseline and federated training under various global learning rates and communication rounds.  The baseline is trained following the traditional centralized recipe, which serves as an iid data oracle for FL with non-iid data. $\eta\!=\!1.0$ indicates a simple model averaging.}\label{tab:fl}
\end{table}

The experimental results of CAFT are given in Table \ref{tab:flcat}. The baseline is trained by first estimating the client-dependent transform and then pooling the canonicalized client data together for model optimization, which is analogous to the cluster adaptive training \cite{Gales_SAT} when dealing with cluster heterogeneity. Compared to the baseline in Table \ref{tab:fl}, the client adaptive training baseline improves the WERs for all the test sets. On average it reduces the WER from 15.83\% to 14.50\%. It is also treated as an oracle model for CAFT. The second block of Table \ref{tab:flcat} shows the WERs under various rounds of communication with a global learning rate of 0.95. The affine transform is $120\!\times\!121$ in dimension and is optimized using SGD with a learning rate of 0.02, a momentum of 0.9, and a batch size of 1,024. We find that a large batch size is helpful when estimating the client-dependent transform. The model is optimized using SGD with a learning rate of 0.05, a momentum of 0.9, and a batch size of 128. For both transform and model optimization, the learning rates anneal by $\frac{1}{\sqrt{2}}$ after the 10th epoch and the training finishes in 20 epochs. From the table, it can be seen that using 10 rounds of communication gives an averaged WER 15.92\%. CAFT can be further improved by pre-training. Instead of starting from a randomized global model across the clients to estimate the client-dependent transform, the global model is initialized with the well-trained FL model without canonicalization. In this case, we use the model from the 8th row of Table \ref{tab:fl} (under $\eta\!=\!0.95$ and $T\!=\!20$ with an averaged WER 16.33\%). The learning rate  anneals by $\frac{1}{\sqrt{2}}$ after the 3rd epoch and the training finishes in 10 epochs. The results are given in the third block of Table \ref{tab:flcat} under CAFT-PT. It can be observed that the pre-training consistently improves the performance on all test sets. With 15 rounds of communication, the averaged WER is 15.13\%, which is about 0.8\% better than without pre-training.

\begin{table}[tbh]
\centering
\begin{tabular}{c c | c c c c c} \hline
      $\eta$    &    $T$         &    S1   &   S2   &   S3   &   S4   &   avg   \\ \hline\hline
   \multicolumn{2}{c|}{\textbf{CAT baseline}} & \textbf{13.0}  &  \textbf{22.4}  &  \textbf{11.5}  &  \textbf{11.1}  &  \textbf{14.50}  \\ \hline
   \multicolumn{7}{c}{\addstackgap{\textbf{CAFT}}}   \\
\textbf{0.95}      &   \textbf{10}         &   \textbf{13.7}  &  \textbf{23.9}  &  \textbf{13.2}  &  \textbf{12.9}  &  \textbf{15.92}  \\
       0.95      &     15         &   13.8  &  25.0  &  13.0  &  12.6  &  16.10  \\
       0.95      &     20         &   13.7  &  25.5  &  12.6  &  12.6  &  16.10  \\
       0.95      &     25         &   13.8  &  24.4  &  12.6  &  12.9  &  15.93  \\ \hline
   \multicolumn{7}{c}{\addstackgap{\textbf{CAFT-PT}}}  \\
       0.95      &     10         &   13.5  &  22.5  &  12.7  &  12.4  &  15.28  \\
\textbf{0.95}    &  \textbf{15}   &   \textbf{13.4}  &  \textbf{22.8}  &   \textbf{12.3}  &   \textbf{12.0}  &   \textbf{15.13}  \\
       0.95      &     20         &   13.4  &  23.6  &  12.1  &  11.9  &  15.25  \\  \hline
\end{tabular}
\caption{WERs of CAT baseline and CAFT under various settings. CATF-PT presents results of CAFT starting with a global pre-trained federated model. The CAT baseline serves as an oracle for CAFT.}\label{tab:flcat}
\end{table}

Table \ref{tab:wgt} shows the WERs of the four test sets under various weighting strategies using CAFT-PT with $\eta\!=\!0.95$ and $T\!=\!15$. The first row uses equal weights on all five clients. In the second row, weights are proportional to the amount of data of each client. Rows 3-5 represent strategies with a preference for a particular client. It can be observed that the ASR performance of a particular client will be improved in most cases if its local model update is favored in weighting in training.

\begin{table}[tbh]
\centering
\begin{tabular}{c | c c c c } \hline
       weight                                                &    S1   &   S2   &   S3   &   S4    \\ \hline\hline
   $p_{i}\!=\!1/L$                                           &  13.4   &  22.8  &  12.3  &  12.0   \\
   $p_{i}\!=\!n_{i}/n$                                       &  13.5   &  24.7  &  14.7  &  14.2   \\
   $p_{\text{accented}}\!=\!0.4$, $p_{i}\!=\!0.15$ rest      &  13.5   &  22.6  &  11.8  &  11.4   \\
   $p_{\text{brodcastnews}}\!=\!0.4$, $p_{i}\!=\!0.15$ rest  &  13.4   &  23.2  &  12.8  &  12.5   \\
   $p_{\text{hospitality}}\!=\!0.4$, $p_{i}\!=\!0.15$ rest   &  13.5   &  22.0  &  12.4  &  11.9   \\  \hline
\end{tabular}
\caption{WERs of CAFT with pre-training under various weighting strategies.}\label{tab:wgt}
\end{table}

\section{Discussion}
\label{sec:dis}

Cross-silo federated training of acoustic models is distributed by design under fundamental requirements of local data storage and communication only permissible between clients and the service provider. Other than those requirements, the local clients have the flexibility of choosing the training strategy to update the local model. For instance, the local $K$-step optimization can be realized in a distributed fashion by itself across multiple nodes whether synchronously or asynchronously. In addition, in this work we assume the communication between clients and the service provider is secure. Otherwise channel encryption techniques or differential privacy \cite{Dwork_DP,Naseri_DP} may be applied to ensure data safety in the training.

The investigated federated training strategy includes a special case where there is only one client and the service provider wants to collaboratively train the model with some public data or its own internal data. This is a common customized acoustic modeling scenario in the real world. In this case, $L=2$ and the second client is the service provider itself while in the mean time its local server is the central server.

Despite the use of an affine transform in this work, in general the client-dependent transform $\mathcal{F}$ can be any nonlinear function. At test time, we use the transform estimated from the training data as we assume the training and test data is matched for a given client. The transform can be further adapted on the test sets, but this estimation is typically unsupervised and needs a two-pass decoding, which will increase inference complexity and introduce latency.

Lastly, it is fairly common that the data from multiple clients is unbalanced. Strategic weighting in training is one way to deal with unbalanced data from clients or if the service provider has certain performance priority in mind.

\section{Summary}
\label{sec:sum}

In this paper we investigated cross-silo federated acoustic modeling to protect data privacy where the ASR service provider collaboratively trains a global acoustic model across multiple clients. Each client has its local data storage which is not shared with either the service provider or the other clients. Models are updated locally from each client before being sent  back to the service provider's central server for aggregation. Various federated averaging schemes have been compared and the impact of the communication frequency is also studied. To deal with the non-iid issue, client adaptive federated training is introduced where a client-dependent transform is estimated to canonicalize the heterogeneous data among the clients. Experiments show that client adaptive federated training can effectively mitigate the data heterogeneity to consistently give improved performance on all the test sets.

\bibliographystyle{IEEEbib}
\bibliography{flcat}

\end{document}